# Emission from Hot Dust in the Infrared Spectra of Gamma-ray Bright Blazars

Michael P. Malmrose[1], Alan P. Marscher[1], Svetlana G. Jorstad[1,2], Robert Nikutta[3], and Moshe Elitzur[3]

## ABSTRACT

A possible source of $\gamma$-ray photons observed from the jets of blazars is inverse Compton scattering by relativistic electrons of infrared seed photons from a hot, dusty torus in the nucleus. We use observations from the Spitzer Space Telescope to search for signatures of such dust in the infrared spectra of four $\gamma$-ray bright blazars, the quasars 4C 21.35, CTA102, and PKS 1510−089, and the BL Lacertae object ON231. The spectral energy distribution (SED) of 4C 21.35 contains a prominent infrared excess indicative of dust emission. After subtracting a non-thermal component with a power-law spectrum, we fit a dust model to the residual SED. The model consists of a blackbody with temperature $\sim$ 1200 K, plus a much weaker optically thin component at $\sim$ 660 K. The total luminosity of the thermal dust emission is $7.9 \pm 0.2 \times 10^{45}$ erg s$^{-1}$. If the dust lies in an equatorial torus, the density of IR photons from the torus is sufficient to explain the $\gamma$-ray flux from 4C 21.35 as long as the scattering occurs within a few parsecs of the central engine. We also report a tentative detection of dust in the quasar CTA102, in which the luminosity of the infrared excess is $7 \pm 2 \times 10^{45}$ erg s$^{-1}$. However, in CTA102 the far-IR spectra are too noisy to detect the 10$\mu$m silicate feature. Upper limits to the luminosity from thermal emission from dust in PKS 1510-089, and ON231, are, $2.3 \times 10^{45}$, and $6.6 \times 10^{43}$ erg s$^{-1}$, respectively. These upper limits do not rule out the possibility of inverse Compton up-scattering of IR photons to $\gamma$-ray energies in these two sources. The estimated covering factor of the hot dust in 4C 21.35, 22%, is similar to that of non-blazar quasars; however, 4C 21.35 is deficient in cooler dust.

*Subject headings:* galaxies: active — infrared: galaxies — quasars: individual (4C 21.35, PKS 1510-089, CTA102)—BL Lacertae Objects: individual(ON231)

[1]Institute for Astrophysical Research, Boston University, 725 Commonwealth Avenue, Boston, MA 02215

[2]Astronomical Institute, St. Petersburg State University, Universitetskij Pr. 28, Petrodvorets, 198504 St. Petersburg, Russia

[3]Department of Physics and Astronomy, University of Kentucky, Lexington, KY 40506-0055

## 1. Introduction

The spectral energy distribution (SED) of a blazar over most of the electromagnetic spectrum is dominated by nonthermal emission from a relativistic jet (see, e.g., Marscher 2010). The continuum at radio through infrared, optical, or even X-ray wavelengths is mainly or completely from synchrotron radiation. The $\gamma$-ray flux, which can dominate the SED much of the time (e.g., von Montigny et al. 1995; Abdo & et al. 2010b), is probably generated by inverse Compton scattering of lower energy seed photons by relativistic electrons in the jet (Sikora et al. 2009).

The source of seed photons for the inverse Compton scattering is controversial. Begelman & Sikora (1987) and Dermer & Schlickeiser (1993) have suggested that photons emitted by the accretion disk, possibly reprocessed by broad emission-line clouds or dust, are scattered to higher energies—the "external-radiation Compton" (ERC) scenario. Multi-frequency monitoring campaigns, on the other hand, indicate that flares in $\gamma$-ray flux often occur after outbursts in the 20-40 GHz range have begun and apparently superluminal knots have appeared in very-long-baseline interferometric images (Jorstad et al. 2001; Lähteenmäki & Valtaoja 2003; Marscher et al. 2010). This implies that the $\gamma$-ray emission has an origin parsecs from the central engine of the active galactic nucleus (AGN). In this case, photons originating from the accretion disk or broad-line region would strike the jet plasma nearly from behind, so that the efficiency of inverse Compton scattering would be too low to explain the high $\gamma$-ray fluxes observed (Sikora et al. 2009).

A variation of the ERC mechanism invokes as the main source of seed photons a molecular torus containing hot dust positioned roughly between $\sim 1$ and 10 pc from the central engine (Błażejowski et al. 2000). Such a torus would be similar to that inferred from the SEDs of quasars (Hao et al. 2005b; Ogle et al. 2006; Cleary et al. 2007) and directly imaged in the Seyfert 2 galaxy NGC 1068 (Jaffe et al. 2004; Raban et al. 2009). Thermal radiation, from warm-to-hot (T $\approx$ 1100 K) dust, has long been thought to be primarily responsible for the infrared (IR) SEDs of radio quiet quasars (Barvainis 1990, 1992). However, detection of a similar dust component in the IR SEDs of blazars has proven difficult owing to the dominance of a nonthermal component. The presence of a hot dust component has been inferred in the quasar 3C 273 (Wills 1989; Soldi et al. 2008), which is a luminous $\gamma$-ray source, based on its SED and variability as a function of wavelength. The case of 3C 273 is, however, difficult to disentangle with certainty because its IR emission includes a variable synchrotron component whose SED peaks near the same wavelength as that of the derived dust feature.

In this paper, we report *Spitzer* Space Telescope (SST, Werner et al. 2004) observations of four $\gamma$-ray bright blazars resulting in the detection of thermal emission from hot dust in the IR spectrum of the quasars 4C 21.35 (PG1222+216), and CTA102, and upper limits to the dust luminosity in the other two objects. In §2 we give an account of the data acquisition, reduction, and analysis, while in §3 we model the SED with a power-law continuum plus radiation from a torus containing hot dust. Section 4 discusses the broader implications of these results, and §5 provides a summary. We adopt a cosmology consisting of a flat universe with $\Omega_\Lambda = 0.73$, $\Omega_m = 0.27$, and $H_0 = 71$ km s$^{-1}$ Mpc$^{-1}$.

## 2. Observations and Analysis

The small sample selected for our study consists of the four blazars detected at $\gamma$-ray energies by both the *Compton* EGRET and *Fermi* LAT instruments (Hartman et al. 1999; Abdo & et al. 2010a) that we found to contain an apparent IR excess over a pure power law in their continuum spectra compiled by Impey & Neugebauer (1988): 4C 21.35 ($z = 0.435$), CTA102 ($z = 1.037$), PKS 1510-089 ($z = 0.361$), and ON 231 ($z = 0.102$). We obtained infrared observations of these blazars with the Infrared Array Camera (IRAC, Fazio et al. 2004), Infrared Spectrograph (IRS Houck et al. 2004), and Multi-Band Imaging Photometer for *Spitzer* (MIPS, Fazio et al. 2004) instruments aboard SST between March 20, 2007 and February 16, 2008. This involved a complete set of observations with each instrument for each target at two different epochs separated by several months. Table 1 presents the observation times and the luminosity distance $D_l$ for each blazar.

Each source was observed with both the Short-Low (SL) and Long-Low (LL) modules of IRS in order to acquire a spectrum over the observed range $\lambda = 5.0\text{-}36.0$ $\mu$m. A mosaiced spectrum of the target was created with observations from each of two slits of the spectrometer. The spectra from the two slits were then averaged together and smoothed over 10 adjacent wavelength bins. The IRAC observations employed either a 5-position Gaussian or 12-position Reuleaux triangle dither pattern, with 2 s frames. Both the MIPS and IRAC photometric measurements were accomplished with the BCD images and Post-BCD products, the latter of which combined all individual BCD frames to form a single mosaiced image at each IRAC (3.6, 4.5, 5.8, and 8.0 $\mu$m) and MIPS (24, 70, and 160 $\mu$m) band. We then applied a correction to account for our finite-sized aperture and a color adjustment to the measured flux according to the SST guidelines contained in the MIPS[4] and IRAC[5] handbooks. The spectral overlap between the IRS and other two SST instruments allowed us to scale the fluxes to compensate for temporal variability or systematic errors in calibration.

In order to cover a wider spectral range, we obtained $R$-band images of each blazar during the epochs of the SST observations with the Perkins Reimaging System (PRISM) on the Perkins 1.8 m telescope of Lowell Observatory in Flagstaff, Arizona. For 4C 21.35, we also include flux densities at five optical bands from the Sloan Digital Sky Survey (SDSS; Schneider et al. 2010), obtained only two days after epoch 2, and at three IR bands from the Two Micron All Sky Survey [6] (2MASS; Skrutskie et al. 2006). The 2MASS magnitudes were converted into flux densities using the zero magnitude flux densities published by Cohen et al. (2003). For comparison, we add later data from the Ultraviolet/Optical Telescope (UVOT) of the *Swift* satellite (Roming et al. 2005), obtained on 16 June 2010 during a major outburst in flux. We calibrated these data following the method of Poole et al. (2008). The optical and UV data were corrected for extinction, according to Schlegel et al. (1998), interpolated or extrapolated at some bands.

---

[4] http://ssc.spitzer.caltech.edu/mips/mipsinstrumenthandbook/51/

[5] http://ssc.spitzer.caltech.edu/irac/iracinstrumenthandbook/101/

Preliminary inspection of the observed far-IR spectrum (i.e., 15-160 $\mu$m) indicates that synchrotron emission originating in the jet dominates this portion of the SED. This is supported by the optical and millimeter-wave linear polarization (Jorstad et al. 2007), and can be described by a power law $F_\lambda \propto \lambda^{-\beta}$ (i.e., $F_\nu \propto \nu^{-\alpha}$, where $\alpha = 2 - \beta$ is the spectral index). To determine the nonthermal portion of the SED, we plot $F_\lambda$ vs. $\lambda$ for each source, with $\lambda$ converted to the AGN rest frame. We then use the MIPS fluxes to fit a power law to the SED by applying the least-squares criterion to the logarithmic plot of $F_\lambda$ vs. $\lambda$. Panels (a) and (b) of Figure 1 display the measured continuum spectra for 4C 21.35, along with the power law that fits the observed 15-160 $\mu$m data. Panels (c) and (d) plot the observed SED, while panels (e) and (f) present the SED after subtraction of the power law.

## 3. Results

The signature of dust components in the SED is an infrared excess with a graybody or blackbody shape after removal of the nonthermal components (Gear et al. 1985). In the case of optically thin silicate dust, prominent features occur in the spectra at $\lambda \approx 10$ $\mu$m and $\lambda \approx 18$ $\mu$m. These are thought to be caused by stretching the Si-O bond and bending of the O-Si-O bond respectively (Knacke & Thomson 1973). In the case of 4C 21.35, the flat IR spectrum strongly suggests the presence of a thermal emitter even before subtraction of the power-law nonthermal component. The power law, with spectral index $\alpha = 0.84$, was derived using the MIPS observations during epoch 1. The same power law was used for epoch 2, when the MIPS 160 $\mu$m observation failed. The variability of flux across epochs was slight, which justifies this approach.

We first fit the residual SED (Fig. 1e,f) with a model consisting of a clumpy, dusty molecular torus (the CLUMPY model of Nenkova et al. 2008a,b), plus a uniform blackbody component with $T \approx 1200$ K. However, this model did not agree with the data as well as a simpler structure consisting of a blackbody of temperature 1200 K and a more weakly emitting optically thin dust component of temperature 660 K. The spectrum of this model, shown in panels (c) and (d) of Figure 1, results in a reduced $\chi^2$ of $\sim 13$. The silicate feature at 10 $\mu$m, apparent in the observed spectrum, appears in emission. The integrated luminosity of the dust emission is $7.9 \pm 0.2 \times 10^{45}$ erg s$^{-1}$. From the luminosity and temperature, we calculate that the surface area of the 1200 K blackbody emitter is equivalent to that of a sphere of radius 2 pc or a torus extending from $\sim 1$ to 2 pc from the central engine. It is possible, however, that the dust is contained in multiple clouds, as is the case for the emission-line gas (see Nenkova et al. 2008b). The blackbody temperature of 1200 K in the model for 4C 21.35 is within the range that causes sublimation of dust (Salpeter 1977), and hence may represent the periphery of dense clouds where sublimation occurs.

---

[6] http://www.ipac.caltech.edu/2mass/releases/allsky/

Equation (1) of Nenkova et al. (2008b) gives the sublimation radius $R_d$ for the dust torus at sublimation temperature $T_{sub}$ as,

$$R_d \simeq 0.4 \left(\frac{L}{10^{45} \text{ erg s}^{-1}}\right)^{1/2} \left(\frac{1500 \text{ K}}{T_{sub}}\right)^{2.6} \text{ pc.} \qquad (1)$$

From this equation we calculate a sublimation radius of $\sim 4$ pc for a $T_{sub}$ of 1200 K. However, as Nenkova et al. (2008b) point out, the reverberation measurements of Minezaki et al. (2004) and Suganuma et al. (2006) yield an inner radius of the dust-emitting regions of AGN that is 2-3 times smaller than the theoretical value, perhaps because only the largest dust grains persist in the inner torus. A torus of hot dust 1-2 pc from the central engine is thus consistent with these studies.

The power laws that best fit the observed 15-160 $\mu$m data for the other two blazars do not result in much infrared excess. We therefore derive only an upper limit to the dust contribution to the SED. This requires subtraction of the steepest power-law that can fit the longer wavelength and R-band data while leaving an IR excess in the residual spectrum. Numerical integration of the residual spectrum over wavelength yields an upper limit to the dust luminosity. Figures 2-4 display the observed spectra, power-law fits, and residual spectra. A closer look at the residual spectra from $\lambda = 5 - 40$ $\mu$m is shown in Figure 5. It is apparent that the IR excess emission over a pure power law in the quasar 4C 21.35 is an order of magnitude greater than in the other three blazars in our sample. In the case of CTA102, extrapolation to the shorter wavelengths of the power-law that fits the spectrum at longer IRS wavelengths yields an excess consistent with dust emission at roughly the same temperature as found for 4C 21.35, $\sim 1200$ K. The luminosity of this excess is $7 \pm 2 \times 10^{45}$ erg s$^{-1}$. The signal is too weak however, to determine whether the 10 $\mu$m feature is present. We therefore consider the detection of dust emission in CTA102 to be tentative. The results for the other two objects are strictly 2-$\sigma$ upper limits: (1) PKS 1510-089: $\alpha = 0.9$, $L_{\text{dust}} < 2.3 \times 10^{45}$ erg s$^{-1}$ (epoch 1); (2) ON231: $\alpha = 1.4$, $L_{\text{dust}} < 6.6 \times 10^{43}$ erg s$^{-1}$ (epoch 1).

## 4. Discussion

### 4.1. Dust Properties of 4C 21.35 and Comparison with Non-blazars

The quasar 4C 21.35 contains a bent, one-sided extended radio jet that ends in a hotspot, with a diffuse, complex lobe on the counterjet side (Jorstad & Marscher 2006). Its radio emission is therefore rather peculiar, although this is partly caused by projection effects owing to a small angle between the jet axis and the line of sight. The deficiency of cool dust in 4C 21.35 is an anomaly relative to AGN that do not exhibit blazar-like behavior. Previously published IR SEDs of radio-loud quasars and radio galaxies with dust emission features (Ogle et al. 2006; Cleary et al. 2007; Schweitzer et al. 2008; Mor et al. 2009), none of which is classified as a blazar, differ strongly from that of 4C 21.35. Siebenmorgen et al. (2005) observed the $10\mu$m silicate feature in emission with IRS in the quasars 3C 249 and 3C 351, and fit photometric data with three components, with the IR emission dominated by $\sim 60$ K dust. Landt et al. (2010) employed all three of *Spitzer*'s science instruments to investigate the dust geometry for a sample of 12 AGN at various orientations. Of that sample, only the Seyfert 1 galaxy, 3C 263, at an orientation angle of $\theta = 34$ °, was found to be deficient in cooler dust. Hao et al. (2005b) observed the 10 $\mu$m and 18 $\mu$m silicate features in five quasars, but were unable to derive reliable temperatures for the dust. In general, it appears that cold dust dominates the IR SEDs of radio-loud quasars where dust emission has been detected. It is therefore striking that emission from cool to warm dust is suppressed in 4C 21.35.

The bolometric luminosity of the central engine of 4C 21.35 is $3.5 \times 10^{46}$ erg s$^{-1}$ if we use the same bolometric conversion as Mor et al. (2009). The dust IR to bolometric luminosity ratio is then 0.23. This conforms with the covering factors of hot dust, 0.20-0.33, derived by Hao et al. (2005a). However, the blazar's 2-25 $\mu$m rest-frame spectrum is considerably flatter, $\alpha = 0.29$, compared with a mean of 1.2 for the quasars in the Ogle et al. study. The drop in flux toward longer wavelengths in the residual spectrum of 4C 21.35 (Fig. 1) contrasts strongly with the nearly constant or rising flux of the non-blazars up to $\gtrsim 20$ $\mu$m. This is true (although not as extreme) even if we do not subtract a power-law nonthermal spectrum (Fig. 1c,d), in the unlikely limit that the entire IR SED is from the thermal dust component at the two observing epochs.

Hence, there is an excess of hot dust at $T \sim 1200$ K and probably a deficit of warm dust at temperatures between $\sim 150$ and 800 K in 4C 21.35 compared with the non-blazars. We note that this contrasts strongly with the finding of Landt et al. (2010) and of (Siebenmorgen et al. 2005) that the warm ($T \sim 300$-800 K) and cold ($T \sim 150$-250 K) dust component is stronger than the hot dust component for sources viewed closer to the jet axis. The hot dust in 4C 21.35 has quite similar properties to that inferred in the $\gamma$-ray bright blazar 3C 273 by Soldi et al. (2008). These authors fit the dust component of the IR emission with a graybody of temperature $\sim 1200$ K with a visible surface area of 1-2 pc$^2$, although uncertain extrapolation of the nonthermal IR component precludes a strong conclusion regarding the relative abundance of dust at lower temperatures.

Our results suggest that a $\gamma$-ray bright blazar is a rare phenomenon, not simply a common radio galaxy with a jet that happens to point almost directly toward our line of sight. We can understand this if emission at MeV and GeV energies dominates the SED of a quasar only if the jet has a high Lorentz factor (for observational evidence of this, see Jorstad et al. 2001; Kellermann et al. 2004; Savolainen et al. 2010), and if such highly relativistic quasars are rare (see Lister & Marscher 1997).

The high hot to cool/warm dust ratio in 4C 21.35 compared with non-blazars implies that the presence of a highly relativistic jet either heats or destroys some of the cooler dust. The jet emission is highly beamed, at least on parsec scales, presumably along the poles of the accretion disk and molecular torus. Measurement of apparent superluminal motion in the jet of 4C 21.35 gives a bulk Lorentz factor $\Gamma \approx 20$ (Jorstad et al. 2001) on parsec scales, with the jet axis pointing within $\sim 3°$ of our line of sight. However, the optical-ultraviolet SED of 4C 21.35 (Fig. 1) indicates that the big blue bump dominates over nonthermal emission from the jet, so that the jet supplies a minority of the photons that could affect the physical state of the dust. Nevertheless, it is possible for photons emitted by the jet to illuminate and heat the upper (i.e., farthest from the equator) layer of a dust torus that is shielded from the central engine by the inner torus, as in the geometry adopted by Kawaguchi & Mori (2010). Given the beaming of the parsec-scale jet emission, this process would be most effective if the bulk Lorentz factor of the section of the jet $\lesssim 1$ pc from the central engine were less than that measured farther downstream (see Marscher et al. 2008). Alternatively, a slower sheath of the jet could produce less well-beamed ultraviolet radiation that shines down on and heats the outer torus.

### 4.2. Relation between Dust and Gamma-ray Emission

The quasar 4C 21.35 is one of only 16 blazars from which the *Fermi* LAT[7] detected $\gamma$-ray emission at photon energies of 0.1–300 GeV above $1 \times 10^{-6}$ photons cm$^{-2}$ s$^{-1}$ (with an integration time of 1 week) between 2008 August and 2010 October. The radiation from hot dust provides a high density of seed photons for inverse Compton scattering by relativistic electrons in the jet. Formula (5) of Sokolov & Marscher (2005) gives the ratio of energy density of IR photons from hot dust to that of the magnetic field, in the rest frame of the jet plasma, as

$$u_{\rm phot}/u_{\rm B} \approx 20(\frac{T}{1200~K})^4(\frac{\Gamma}{20})^2 B^{-2}, \tag{2}$$

---

[7]Light curves are displayed at http://fermi.gsfc.nasa.gov/ssc/data/access/lat/msl_lc/

independent of the distance $r_{\rm in}$ from the central engine. Theoretically, this is roughly the same as the ratio of ERC to synchrotron luminosity. Here, $T$ is the blackbody temperature of the dust and $B$ is the magnetic field in the jet in gauss. Since the magnetic field strength of a blazar on parsec scales is typically 0.1-1 G, ERC scattering of IR emission from dust dominates over that of synchrotron emission from the jet (Sikora et al. 2009). While it is possible that local sources of seed photons, such as synchrotron emission from a slower sheath of the jet, might compete with the dust radiation (Marscher et al. 2010), the high ratio of $\gamma$-ray to IR synchrotron luminosity observed when 4C 21.35 is in a high state (e.g., during the first half of 2010; Donato 2010) can be explained by inverse Compton scattering of IR photons from the hot dust. Recent analysis by Tanaka et al. (2011) of a GeV outburst in 4C 21.35 observed by the *Fermi* LAT, coincident with a flare at photon energies up to 0.4 TeV observed by the MAGIC Cerenkov telescope (Mariotti 2010), supports the conclusion that the $\gamma$-rays were emitted at distances $\gtrsim 1$ pc from the central engine. This is consistent with parsec-scale hot dust as the source of the seed photons.

Because we obtain a rather high upper limit of 0.25 to to the dust-to-bolometric luminosity ratio for PKS 1510−089 (with bolometric luminosities estimated from the SEDs displayed in the NASA/IPAC Extragalactic Database), and have no constraint on the ratio in the BL Lac object ON231, we cannot exclude the possibility that all $\gamma$-ray bright blazars contain nuclear dust with covering factors exceeding 25%. It is therefore possible that a high flux of IR emission from dust might be required to produce $\gamma$-ray fluxes greater than $1 \times 10^{-6}$ photons cm$^{-2}$ s$^{-1}$ that are sometimes observed in quasars (e.g., Abdo & et al. 2010a). Unfortunately, detection of emission from dust is difficult owing to the dominance of the nonthermal IR component in most blazars. We note, on the other hand, that Leipski et al. (2009), in a study of 15 FR I galaxies with detected optical activity in the nucleus, a significant IR excess is detected in 8 sources (53% of the sample.) In four of these sources the infrared excecss is attributed to warm dust (hundreds of Kelvins) emission powered by a hidden AGN. This implies that the number of BL Lac objects with hot dust ($T > 1000$ K) emission is quite low, since they are thought to reside in FR I galaxies (Urry & Padovani 1995).

## 5. Summary

We have searched for signs of thermal emission from hot dust in the IR spectra of four $\gamma$-ray bright blazars. We find clear evidence for dust emission in the quasar 4C 21.35, whose IR spectrum is nearly flat at rest wavelengths between 2 and 25 $\mu$m. The luminosity of the thermal emission, $\sim 8 \times 10^{45}$ erg s$^{-1}$, is sufficient for the radiation to supply the bulk of the seed photons for inverse Compton scattering to $\gamma$-ray energies during the outburst in the first half of 2010. We also find an infrared excess in the IR spectrum of CTA 102 with a luminosity of $\sim 7 \times 10^{45}$ erg s$^{-1}$ that is consistent with dust emission. The case for dust emission is weak for the other two blazars, for which we can only determine an upper limit after subtraction of the non-thermal component. These upper limits, however, do not exclude inverse Compton scattering of photons emitted by hot dust as the main source of $\gamma$-ray seed photons.

The IR spectrum of 4C 21.35 is considerably flatter than that of other non-blazar quasars, with most of the luminosity arising from the hottest dust. The deficit of dust with temperatures in the 150-800 K range may be related to the presence of the highly relativitistic jet, which could produce ultraviolet synchrotron photons that heat dust in the outer torus, where only warm or cold dust resides in non-blazar active galactic nuclei.

Thermal emission from dust is difficult to find in blazars owing to the strong nonthermal continuum. Except in relatively rare cases such as 4C 21.35, in which the power-law spectrum from synchrotron radiation is less prominent at near- and mid-IR wavelengths, accurate determination and subtraction of the nonthermal component is essential in the search for dust emission. This requires simultaneous flux measurements at many wavelengths between the optical and millimeter spectral regions. In addition, the nonthermal emission is generally variable and linearly polarized, so determination of the SED at different times and measurement of polarization at different IR to millimeter wavelengths can aid in its subtraction (see Soldi et al. 2008). Future multi-wavelength monitoring campaigns would allow a better definition of the properties of the dust torus and how these depend on the nonthermal luminosity and relativistic beaming of the jet.

Clavel et al. (1989), Minezaki et al. (2004), and Suganuma et al. (2006) have observed changes in the thermal IR flux of Seyfert galaxies, with delay times of $\lesssim 1$ year relative to UV variations in the big blue bump emission. Barvainis (1992) interprets this as the consequence of "reverberation" from the heating of silicate or graphite dust to temperatures $T \sim 1100 - 1900$ K. This yields a nearly blackbody spectrum whose variation of total luminosity vs. time is well correlated with changes in the UV luminosity. A model was developed, in which higher UV flux from the central engine increases the radius of the hot-dust region and therefore increases the flux of IR photons emitted from that region. The AGN unification scheme (see the reviews by Antonucci 1993; Urry & Padovani 1995) suggests that similar reverberations in the IR SED of blazars containing dust should be observable. In fact, we expect this to occur over the next few years in 4C 21.35, which exhibited a major outburst in UV flux in 2010 (see Fig. 1). The time delay between the peak of the UV flare and the IR flux from the dust reverberation would then provide valuable information on the location and geometry of the dusty region. Furthermore, any outbursts observed in the UV emission from the big blue bump in other blazars could increase the IR emission to facilitate detection of the dust component.

We thank the referee, R. Antonucci, who provided insightful comments. This work is based on observations made with the *Spitzer* Space Telescope, which is operated by the Jet Propulsion Laboratory (JPL), California Institute of Technology under a contract with NASA. Funding for this project was supplied by JPL through *Spitzer* grant 1289764, by NASA through *Fermi* Guest Investigator grant NNX08AV65G, and by National Science Foundation (NSF) grant AST-0907893. This study included observations with the PRISM camera (Kenneth Janes, P.I.), developed at Boston University and Lowell Observatory, with financial support from NSF, Boston University, and Lowell Observatory. This publication makes use of data products from the Two Micron All Sky Survey, which is a joint project of the University of Massachusetts and the Infrared Processing and Analysis Center/California Institute of Technology, funded by NASA and the NSF. Funding for the SDSS and SDSS-II has been provided by the Alfred P. Sloan Foundation, the Participating Institutions, NSF, the U.S. Department of Energy, NASA, the Japanese Monbukagakusho, the Max Planck Society, and the Higher Education Funding Council for England. The SDSS Web Site is http://www.sdss.org/.

Table 1. Luminosity Distances and Dates of Observations.

| Source | $D_l$ (Gpc) | Epoch | IRS Dates (y-m-d) | MIPS Dates (y-m-d) | IRAC Dates (y-m-d) |
|---|---|---|---|---|---|
| 4C 21.35 | 2.39 | 1 | 2007-06-11 | 2007-07-13 | 2007-06-28 |
|  |  | 2 | 2008-01-12 | 2008-02-16 | 2008-02-01 |
| CTA102 | 6.94 | 1 | 2007-06-27 | 2007-07-13 | 2007-07-02 |
|  |  | 2 | 2007-12-06 | 2008-01-08 | 2007-12-22 |
| PKS 1510-089 | 1.91 | 1 | 2007-03-20 | 2007-03-01 | 2007-03-31 |
|  |  | 2 | 2007-08-06 | 2007-09-15 | 2007-08-07 |
| ON231 | 0.465 | 1 | 2007-06-15 | 2007-06-05 | 2007-07-04 |
|  |  | 2 | 2008-01-10 | 2008-01-08 | 2008-01-30 |

Table 2 (a).    Data for 4C 21.35

| | Epoch 1 | | | Epoch 2 | |
|---|---|---|---|---|---|
| $\lambda_{rest}$ | $F_\nu$ | $\sigma F_\nu$ | $\lambda_{rest}$ | $F_\nu$ | $\sigma F_\nu$ |
| ($\mu$m) | (mJy) | (mJy) | ($\mu$m) | (mJy) | (mJy) |
| *IRS* | | | | | |
| 3.57 | 2.42 | 0.87 | 3.59 | 12.75 | 1.88 |
| 3.59 | 11.21 | 1.80 | 3.61 | 14.18 | 0.84 |
| 3.61 | 11.82 | 0.87 | 3.63 | 14.45 | 0.53 |
| *IRAC* | | | | | |
| 2.47 | 8.36 | 1.10 | 2.47 | 8.61 | 0.28 |
| 3.13 | 10.82 | 1.25 | 3.13 | 11.11 | 0.32 |
| 3.99 | 12.50 | 1.36 | 3.99 | 12.85 | 0.35 |
| 5.42 | 12.71 | 1.38 | 5.42 | 13.31 | 0.35 |
| *MIPS* | | | | | |
| 16.72 | 21.03 | 1.87 | 16.72 | 21.59 | 1.90 |
| 48.78 | 81.32 | 8.59 | 48.78 | 88.76 | 8.52 |
| 111.50 | 159.80 | 35.99 | - | - | - |

Note. — Only the first three lines of the unsmoothed IRS data are presented. The entire contents of Table 2 (a)-(d) are provided as a machine readable table in the online journal.

Table 2 (b).   Data for PKS 1510-089

| $\lambda_{rest}$ ($\mu$m) | Epoch 1 $F_\nu$ (mJy) | $\sigma F_\nu$ (mJy) | $\lambda_{rest}$ ($\mu$m) | Epoch 2 $F_\nu$ (mJy) | $\sigma F_\nu$ (mJy) |
|---|---|---|---|---|---|
| *IRS* | | | | | |
| 3.76 | 0.002.43 | 1.17 | 3.76 | 4.12 | 1.22 |
| 3.78 | 0.011.57 | 2.14 | 3.78 | 15.32 | 3.98 |
| 3.81 | 0.012.16 | 1.00 | 3.81 | 15.42 | 4.20 |
| *IRAC* | | | | | |
| 2.61 | 8.51 | 0.28 | 2.61 | 12.11 | 0.33 |
| 3.30 | 10.86 | 0.31 | 3.30 | 15.67 | 0.38 |
| 4.21 | 12.98 | 0.35 | 4.21 | 19.72 | 0.43 |
| 5.72 | 16.20 | 0.39 | 5.72 | 24.05 | 0.48 |
| *MIPS* | | | | | |
| 17.63 | 42.10 | 2.64 | 17.63 | 71.73 | 3.30 |
| 51.43 | 96.51 | 10.28 | 51.43 | 157.43 | 12.04 |
| 117.56 | 190.03 | 46.07 | 117.56 | 251.21 | 46.07 |

Note. — Only the first three lines of the unsmoothed IRS data are presented. The entire contents of Table 2 (a)-(d) are provided as a machine readable table in the online journal.

Table 2 (c).   Data for CTA102

| $\lambda_{rest}$ ($\mu$m) | Epoch 1 $F_\nu$ (mJy) | $\sigma F_\nu$ (mJy) | $\lambda_{rest}$ ($\mu$m) | Epoch 2 $F_\nu$ (mJy) | $\sigma F_\nu$ (mJy) |
|---|---|---|---|---|---|
| *IRS* | | | | | |
| 2.51 | 1.33 | 1.29 | 2.51 | 0.46 | 1.62 |
| 2.53 | 5.83 | 2.35 | 2.53 | 3.79 | 4.63 |
| 2.54 | 8.18 | 1.12 | 2.54 | 4.20 | 1.03 |
| *IRAC* | | | | | |
| 1.74 | 0.33 | 0.17 | 1.74 | 2.30 | 0.14 |
| 2.21 | 0.38 | 0.21 | 2.21 | 3.07 | 0.17 |
| 2.81 | 0.43 | 0.27 | 2.81 | 4.53 | 0.21 |
| 3.82 | 0.48 | 0.32 | 3.82 | 6.29 | 0.25 |
| *MIPS* | | | | | |
| 11.78 | 37.00 | 2.48 | 11.78 | 36.59 | 2.46 |
| 34.36 | 172.16 | 12.12 | 34.36 | 158.73 | 11.89 |
| 78.55 | 280.00 | 43.19 | 78.55 | 225.30 | 44.63 |

Note. — Only the first three lines of the unsmoothed IRS data are presented. The entire contents of Table 2 (a)-(d) are provided as a machine readable table in the online journal.

Table 2 (d).    Data for ON231

| | Epoch 1 | | | Epoch 2 | |
|---|---|---|---|---|---|
| $\lambda_{rest}$ | $F_\nu$ | $\sigma F_\nu$ | $\lambda_{rest}$ | $F_\nu$ | $\sigma F_\nu$ |
| ($\mu$m) | (mJy) | (mJy) | ($\mu$m) | (mJy) | (mJy) |
| *IRS* | | | | | |
| 4.65 | 6.37 | 1.21 | 4.65 | 7.13 | 1.11 |
| 4.67 | 24.44 | 2.21 | 4.67 | 29.56 | 2.12 |
| 4.70 | 24.99 | 1.02 | 4.70 | 31.65 | 1.05 |
| *IRAC* | | | | | |
| 3.22 | 21.06 | 0.44 | 3.22 | 25.38 | 0.48 |
| 4.08 | 24.42 | 0.47 | 4.08 | 29.23 | 0.52 |
| 5.20 | 27.90 | 0.51 | 5.20 | 33.56 | 0.55 |
| 7.06 | 31.90 | 0.55 | 7.06 | 37.44 | 0.60 |
| *MIPS* | | | | | |
| 21.78 | 50.45 | 2.89 | 21.78 | 78.50 | 3.60 |
| 63.52 | 105.18 | 9.82 | 63.52 | 149.75 | 11.28 |
| 145.19 | 212.34 | 41.75 | 145.19 | 230.34 | 41.75 |

Note. — Only the first three lines of the unsmoothed IRS data are presented. The entire contents of Table 2 (a)-(d) are provided as a machine readable table in the online journal.

– 19 –

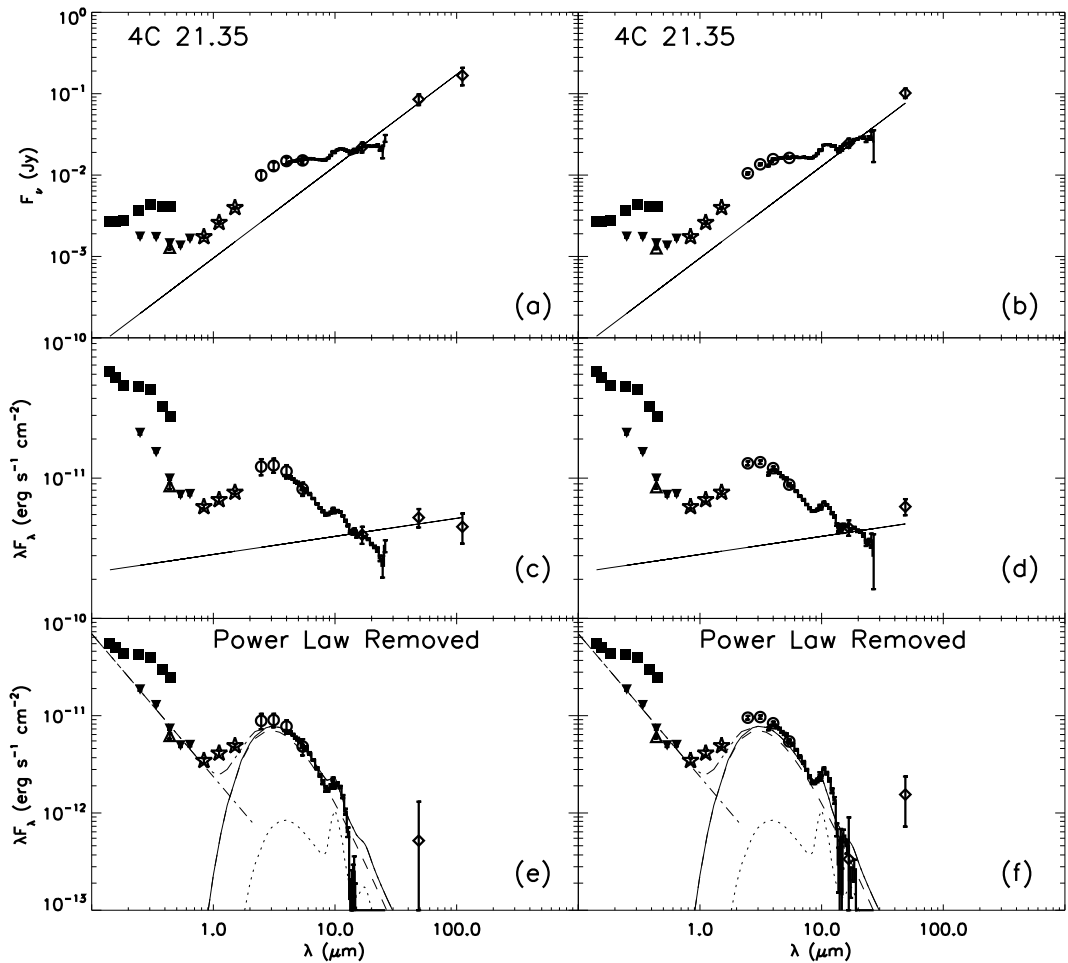

Fig. 1.— (a) and (b): IR to UV spectra of 4C 21.35 ($z = 0.435$) at epochs 1 and 2, respectively. *Horizontal axis*: wavelength in quasar rest frame. *Solid line*: power law (spectral index $\alpha = 0.84$) that best fits measurements at observed wavelengths of 20-160 $\mu$m. *Circles, small squares, and diamonds*: IRAC, IRS, and MIPS data, respectively; *upward pointing triangle*: ground-based R-band observation; stars: fluxes from 2MASS survey; *downward pointing triangles*: SDSS measurements; *large squares*: *Swift* observations during an outburst in June 2010, showing that the optical-UV thermal emission is time-variable. (c) and (d): Observed SED. *Line*: as in panels (a) and (b). (e) and (f): Residual SED after subtraction of power law. *Dashed curve*: SED of a $\sim$ 1200 K blackbody; *dotted curve*: SED of optically thin dust at a temperature of 660 K; solid curve: combination of the two dust components; *dash-dot line*: power-law fit to optical; dash-double-dot curve: sum of all components at near-IR wavelengths.

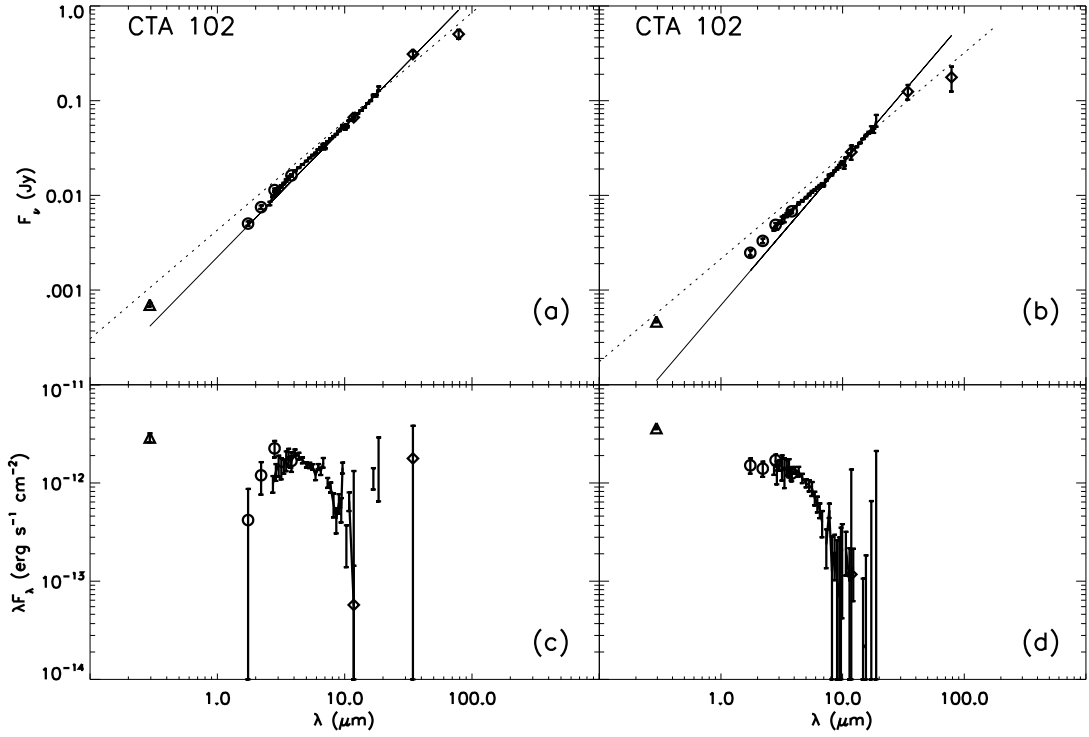

Fig. 2.— (a) and (b) SED of CTA102 ($z = 1.037$) at epochs 1 and 2, respectively. *Dotted line*: power law (spectral index $\alpha = 0.75, 0.79$ for a, b, respectively) that best fits the SED, with spectral index $\alpha = 0.75$ in (a) and $\alpha = 0.79$ in (b). *Solid line*: power law derived from the observations at longer wavelengths of the IRS spectrum, with $\alpha = 0.50$. The residuals after subtraction of the second power law are displayed in panels (c) and (d), the second of which is used to estimate an upper bound on the dust luminosity.

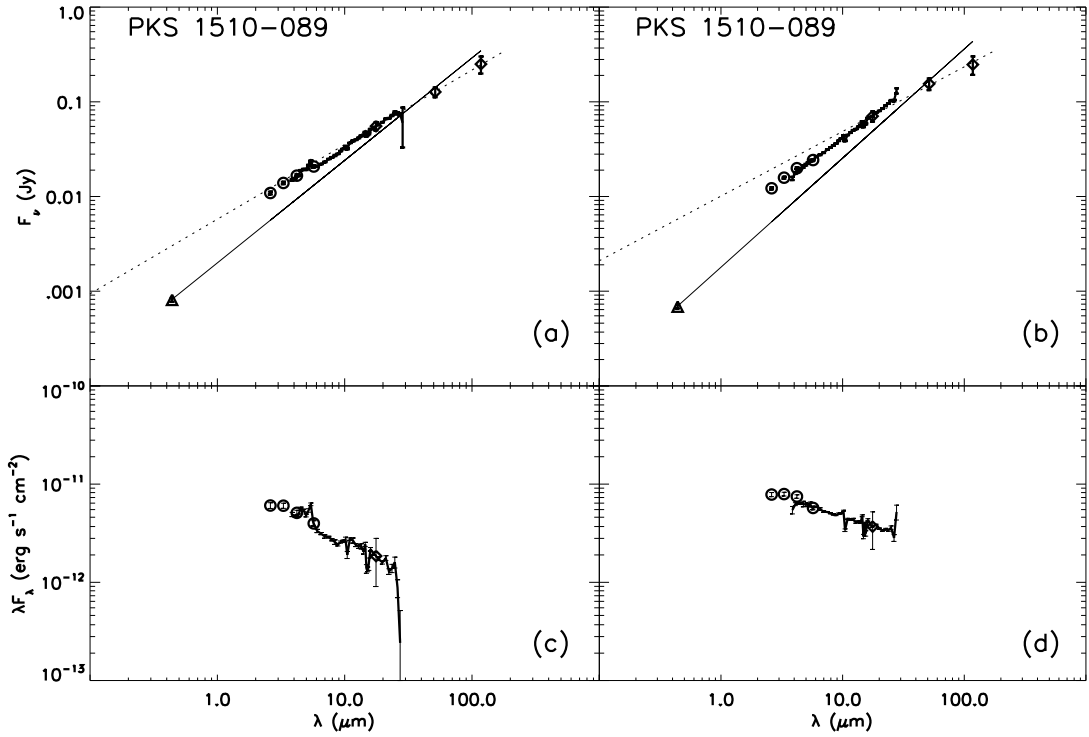

Fig. 3.— As in Fig. 2, for PKS 1510-089. *Dotted line*: power-law derived solely from the MIPS data, with $\alpha = 1.22$, $\alpha = 1.29$ for (a), (b), respectively. *Solid line*: power-law with steepest slope consistent with MIPS and $R$-band data, with $\alpha = 0.91, 0.82$ for (a), (b), respectively. The resultant spectrum in panel (c) is used to derive an upper limit to the luminosity from dust emission.

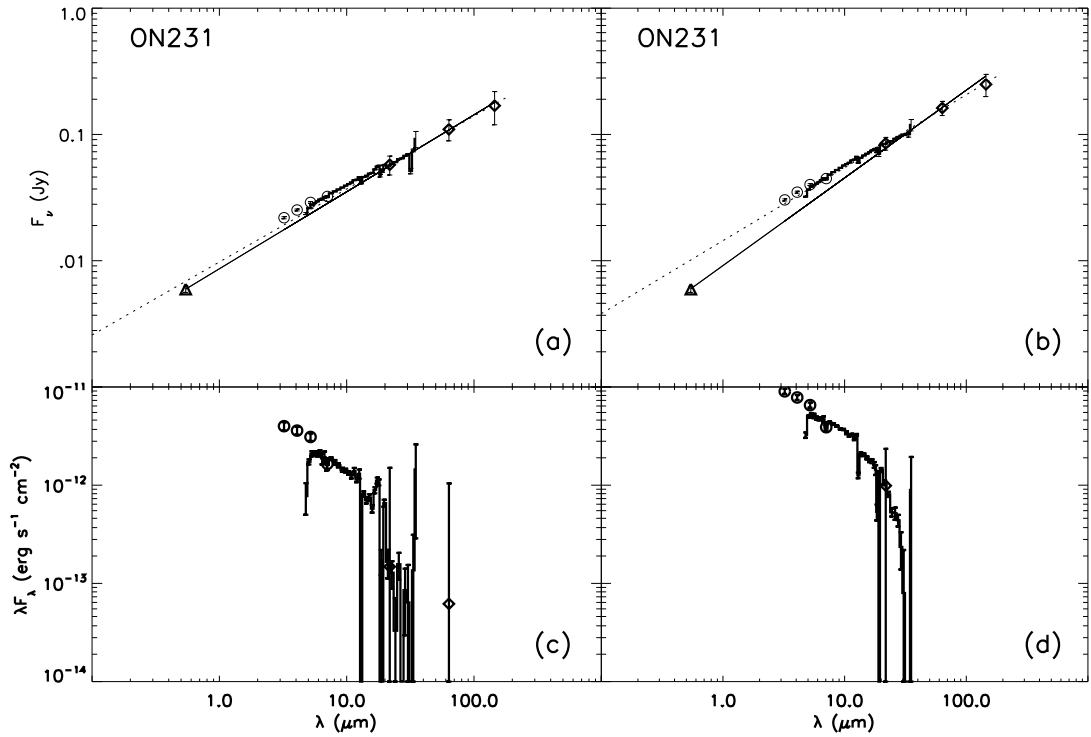

Fig. 4.— As in Fig. 3, for ON231. *Dotted line* corresponds to spectral index $\alpha = 1.41$ at both epochs; *solid line* corresponds to $\alpha = 1.39, 1.30$ for (a), (b), respectively. The residual spectrum in panel (c) provides an upper limit to the luminosity from dust emission.

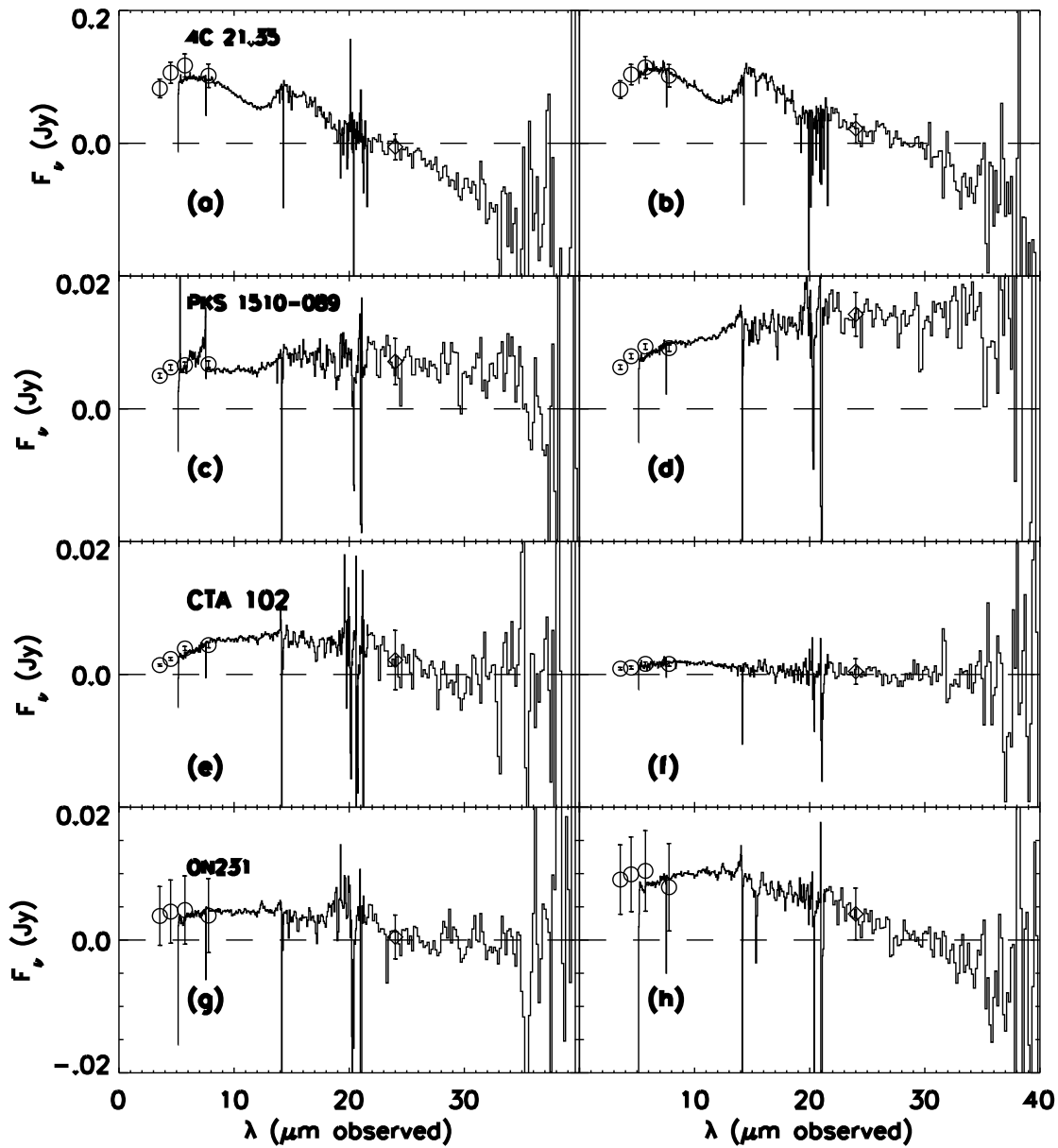

Fig. 5.— A close-up view of the observed spectra of all four blazars at epoch 1 (a, c, e, and g) and epoch 2 (b, d, f, and h), respectively, at $\lambda = 0 - 40$ $\mu$m. Symbols as in Fig. 4. (a-b) 4C 21.35, (c-d) PKS 1510-089, (e-f) CTA 102, (g-h)ON231. *Horizontal axis*: observed wavelengths. *Vertical axis*: flux density in Janskys after the power-law has been removed. Note that the infrared excess in 4C 21.35 is an order of magnitude greater than in the other three blazars. The feature near 14 $\mu$m in all eight spectra is an artifact of the IRS where the instrument observes the shorter wavelengths in the SL module and the longer wavelengths in the LL module.